\documentclass[9pt,twocolumn,twoside,lineno]{pnas-new}
\usepackage{amssymb,amsfonts,amsmath,bm}
\usepackage{subcaption}
\usepackage{tikz}
\usetikzlibrary{positioning}

\templatetype{pnasresearcharticle} 
\newcommand{\tensor}[1]{{\bf\bm{#1}}}
\newcommand{\grad}{\vec{\nabla}}
\newcommand{\tensoriden}{\tensor{\bm{I}}}
\newcommand{\tensorsigma}{\tensor{\bm{\sigma}}}
\newcommand{\tensorsigmanod}{\tensor{\bm{\sigma}_0}}
\newcommand{\change}{}
\DeclareMathOperator{\Tr}{tr}

\title{Multi-scale poromechanics of wet cement paste}

\author[a]{Tingtao Zhou}
\author[b,c,d,1]{Katerina Ioannidou}
\author[d]{Franz-Josef Ulm}
\author[e]{Martin Z. Bazant}
\author[c,d,1]{Roland J.M. Pellenq}

\affil[a]{Massachusetts Institute of Technology, Department of Physics, Cambridge, MA 02139}
\affil[b]{Laboratoire de Mécanique et Génie Civil, CNRS - Université de Montpellier,  Montpellier, France}
\affil[c]{The MIT / CNRS / Aix-Marseille University Joint Laboratory, "Multi-Scale Materials Science for Energy and Environment"}
\affil[d]{Massachusetts Institute of Technology, Department of Civil and Environmental Engineering, Cambridge, MA 02139}
\affil[e]{Massachusetts Institute of Technology, Departments of Mathematics and Chemical Engineering, Cambridge, MA 02139}

\significancestatement{Capillary effects in cement paste are associated with multiple degradation mechanisms. Using a novel framework, we investigated the role of capillary forces in cement paste under partial saturation, from nano-grains level to the mesoscale.
We show that the largest capillary forces concentrate at the boundary between gel pores and larger capillary pores inducing \change{nonaffine} deformations with correlations up to a few tens of nm. Our results suggest a homogenization length scale common to liquid and solid stresses, correlated with the scale of inherent structural heterogeneities. This opens the route to studying the poromechanics of complex multiscale media with explicit fluid-solid coupling. }

\authorcontributions{T.Z., K.I. and R.P. designed research, T.Z. conducted research, T.Z., K.I., F. U., M.B. and R.P. wrote the paper.}
\authordeclaration{The authors declare no conflict of interest.}

\correspondingauthor{\textsuperscript{1}To whom correspondence should be addressed. E-mail: pellenq@mit.edu, aikaterini.ioannidou@umontpellier.fr}


\begin{abstract}
Capillary effects such as imbibition-drying cycles impact the mechanics of granular systems over time. A multiscale poromechanics framework was applied to cement paste, that is the most common building material, experiencing broad humidity variations over the lifetime of infrastructure. First, the liquid density distribution at intermediate to high relative humidities is obtained using a lattice gas density functional method together with a realistic nano-granular model of cement hydrates. The calculated adsorption/desorption isotherms and pore size distributions are discussed and compare well to nitrogen and water experiments. The standard method for pore size distribution determination from desorption data is evaluated. Then, the integration of the Korteweg liquid stress field around each cement hydrate particle provided the capillary forces at the nanoscale. The cement mesoscale structure was relaxed under the action of the capillary forces. Local \change{irreversible} deformations of the cement nano-grains assembly were identified due to liquid-solid interactions. The spatial correlations of the nonaffine displacements extend to a few tens of nm. Finally, the Love-Weber method provided the homogenized liquid stress at the micronscale. The homogenization length coincided with the spatial correlation length of \change{nonaffine displacements}. Our results on the solid response to capillary stress field suggest that the micronscale texture is not affected by mild drying, while nanoscale \change{irreversible} deformations still occur. These results pave the way towards understanding capillary phenomena induced stresses in heterogeneous porous media ranging from construction materials, hydrogels to living systems.

\end{abstract}

\dates{This manuscript was compiled on \today}
\doi{\url{www.pnas.org/cgi/doi/10.1073/pnas.XXXXXXXXXX}}

\begin{document}

\verticaladjustment{-2pt}

\maketitle
\thispagestyle{firststyle}
\ifthenelse{\boolean{shortarticle}}{\ifthenelse{\boolean{singlecolumn}}{\abscontentformatted}{\abscontent}}{}

\dropcap{O}ver the life-time of cement paste, the degree of water saturation can span a wide range: the initial reaction and precipitation of cement hydrates happens when cement powder is mixed with water to form the cement paste. Later, during setting, remaining water (not used by the dissolution-precipitation reaction) gradually evaporates from the pore space \citep{schiessl2000assessing} that is composed of gel pores ($\sim$1-5~nm) and capillary pores (5-50~nm). The set paste in construction environment is then exposed to weather conditions that correspond to different relative humidities. These changes in saturation level of cement paste induce capillary stress and material deformation, as demonstrated in drying shrinkage experiments \citep{feldman1968model, feldman1968model, thomas2008structural, maruyama2015bimodal}, which potentially contributes to the degradation and failure of buildings and bridges.

Structural changes due to liquid intake/drainage in porous media are known as ``sorption induced deformation'' \citep{gor2017adsorption}. Irreversible deformations during drying shrinkage of cement paste have been observed. Under mild drying conditions, small angle neutron scattering (SANS) experiments \citep{thomas2008structural} show plastic rearrangements at very small length scales, but no significant structural changes are detected at larger scales.  However, harsh drying conditions can lead to large strain irreversibility at the macroscale\citep{feldman1968model, maruyama2015bimodal}. It is also known that pressurization of the pore fluid in liquid saturated rocks is an important weakening mechanism that can lead to fracture \citep{brantut2011pore}.

Capillary condensation and induced mechanical strain in materials consisting of simple pore structures (such as MCM-type mesoporous silica and carbon nanotubes) have been successfully described using independent pores of cylindrical geometry\citep{gelb1999phase, pellenq2009simple, gor2010adsorption, gor2017adsorption}. However, the multi-scale pore structure of a heterogeneous material such as cement paste invalidates these theories mainly for 2 reasons. 1) The pores are connected to form a complex percolating topology, which gives rise to more complicated hysteretic behaviors in the sorption isotherms and liquid distributions. 2) The liquid distribution inside the pores renders a heterogeneous force boundary condition, which along with a highly heterogeneous solid structure, challenges the applicability of classical continuum tools \citep{torquato2013random}.
Thus to study the mechanics of undersaturated cement paste, a multi-scale approach must be undertaken.


Homogenization methods play a central role in bridging different scales to understand how large scale properties emerge from small scale interactions. When the material is heterogeneous and partially saturated, it is necessary to re-examine the emergence of a continuum description. One important issue is how to determine a proper physical homogenization scale\citep{rycroft2009assessing}.
For some cases of dense granular flows, the continuum postulate has been shown valid~\citep{rycroft2009assessing, kamrin2007stochastic}. However, this has not  been tested for poromechanics in the presence of capillary forces.

In this paper, we first simulate the liquid distribution in a heterogeneous mesoscale model of cement paste at partial saturations, using lattice gas density functional theory (DFT).
\change{Fig.1 shows the mesoscale model of cement paste composed of cement hydrates nano-grains ($\sim$ 5~nm) interacting with effective potentials from atomistic simulations\cite{bonnaud2016interaction, pellenq2009realistic, ioannidou2014controlling}.}
Our simulated hysteric adsorption/desorption isotherms of water and nitrogen compare well with experiments.
Our subsequent calculations reveal the nanoscale details of capillary forces, by constructing the Korteweg stress tensor field and integrating over Voronoi cells centered on cement hydrate nano-grains\cite{capillary-langmuir}.
\change{Using the capillary forces, we simulate structural relaxation and a subsequent sorption cycle.} Analysis of the statistics of solid and liquid stress distributions point to the same \change{homogenization length}. We rationalize it as the length scale of structural heterogeneity. To our knowledge this is the first example examining the continuum postulate for unsaturated porous media. Our results provide insights to drying/wetting of cement paste.

\begin{figure}[h]
	\centering
	\includegraphics[width=1.0\linewidth]{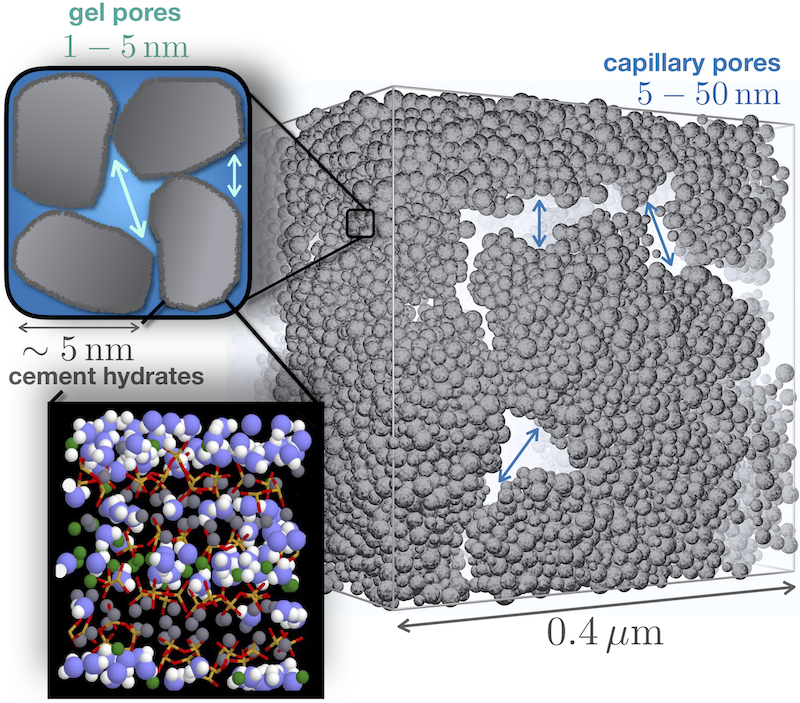}
	\caption{
		\label{fig:cement-structure}
\change{Multiscale cement paste model. Mesoscale structure of a realistic cement paste model spanning a cubic simulation box of $0.4$~$\mu$m. Gray spheres represent nano-grains of cement hydrates of polydisperse diameters $\sim 5$~nm. The upper inset illustration is a zoom-in of few cement hydrates showing the gel pores (1-5~nm) among the closed-packed nano-grains. The molecular structure of cement hydrates in shown in the lower inset (taken from Ref.~\cite{pellenq2009realistic}) - yellow and red sticks are silicon and oxygen atoms in silica tetrahedra, the blue and white spheres are oxygen and hydrogen atoms of water molecules, respectively; the green and gray spheres are inter and intra-layer calcium ions, respectively. The effective pairwise interaction potentials of the nano-grains have been upscaled from atomistic simulations (~\cite{ioannidou2016natcom,bonnaud2016interaction}). The mesoscale cement paste model is based in out-of-equilibirum precipitation and aggregation of cement hydrates (~\cite{ioannidou2014controlling}, see SI Appendix). It reproduces realistically the mesoscale structure and mechanics of cement paste including the complex morphology of capillary (5-50~nm) and gel pores ~\cite{ioannidou2016mesoscale}.}
	}
\end{figure}

\subsection*{Sorption hysteresis and pore size distributions}
\begin{figure*}[b]
	\centering
	\includegraphics[width=1.0\linewidth]{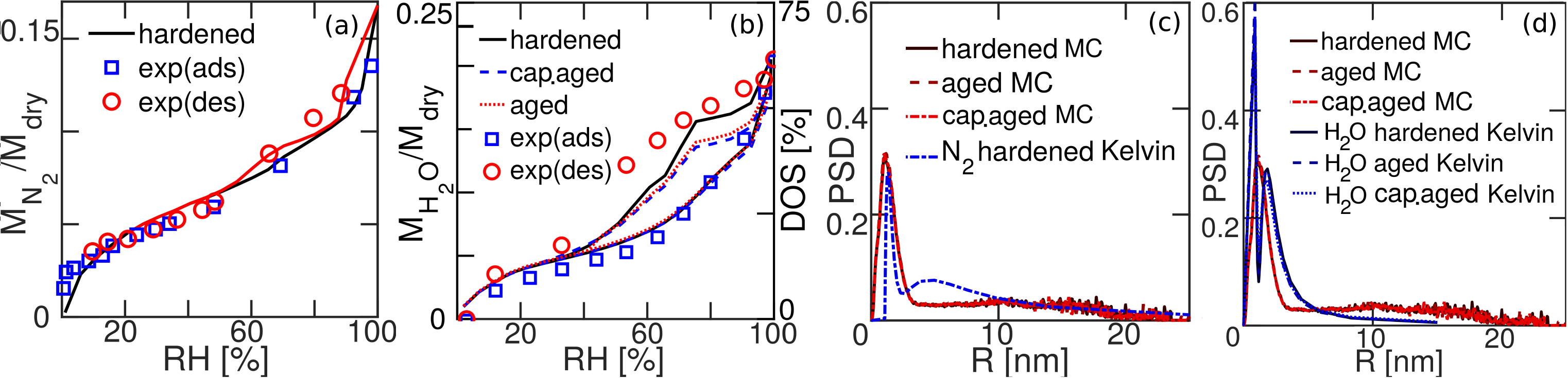}
	\caption{
		\label{fig:sorption-isotherm-psd}
		Isotherms and pore size distributions.
		Adsorption/desorption isotherms for (a) nitrogen (at 77~K) and (b) water (at 300~K) in cement paste . Symbols show the mass ratio of wet over dry sample ($M_{adsorbent}/M_{dry}$ (blue squares and red dots for adsorption and desorption respectively) from experiments of Ref.~\citep{mikhail1964pore} for (a) and Ref.~\citep{baroghel2007water} for (b). Black lines show simulated isotherms on cement model configurations with realistic 3D pore network of water to cement ratio (w/c) $0.52$. For water sorption (b) the degree of saturation (DOS) estimated from the simulations is shown at the right y axis. (c) and (d) show pore size distributions calculated by Monte Carlo method on the three cement model configuration, and from applying the Kelvin's equation to the isotherms of nitrogen and water respectively. Y and X-axis values are normalized probability density and pore radius respectively.
	}
\end{figure*}

To obtain the liquid distributions in cement paste, we simulated the adsorption/desorption isotherms for two adsorbents nitrogen and water in mesoscale configurations of cement hydrates using lattice gas DFT~\citep{kierlik2001capillary, kierlik2002adsorption} described in the Materials and Methods section. The sorption isotherms calculated for both adsorbents are in good agreement with experimental data, as shown in Fig.\ref{fig:sorption-isotherm-psd}a and b for nitrogen (77~K) and water (300~K) respectively. The nitrogen adsorption/desorption isotherm displays only minimal hysteresis in agreement to experimental observations of Ref.\citep{mikhail1964pore}. The mesoscale configuration of cement hydrates used for the nitrogen and water lattice DFT simulation is called ``hardened'' as it is obtained right after out-of-equilibrium precipitation of nano-grains \cite{ioannidou2014controlling} and has internal eigenstresses \citep{ioannidou2017inhomogeneity} (see Methods and SI Appendix).

The water adsorption/desorption isotherm features a significant hysteresis. We focus our analysis to pure capillary effects at $RH$~>$30\%$ and the results compare well with experimental data of Ref.~\cite{baroghel2007water}. Under harsh drying conditions other chemo-physical processes may take place and these are beyond our scope~\citep{bonnaud2012thermodynamics,pinson2015hysteresis}. Our approach allows us to explain the water hysteresis loop from meta-stable ink-bottleneck states upon desorption due to the pore constrictions in cement paste. This contrasts with previous results based on equilibrium thermodynamics for both adsorption and desorption branches and pore connectivity assumption \citep{pinson2015hysteresis, pinson2018inferring}.

To investigate the poromechanical effect of internal eigenstresses and water capillary forces, we simulated a second cycle of adsorption/desorption 1) on the ``aged''  (the hardened configuration relaxed to zero average eigenstress by means of Molecular Dynamics (MD) simulation in the NPT ensemble) and 2) on the ``capillary aged'' (the hardened configuration relaxed to zero average eigenstress by MD simulation in the NPT ensemble under the action of capillary forces at $RH=31\%$ where capillary forces are the largest; see next sections). All curves for the adsorption branch of water are superposable (Fig.\ref{fig:sorption-isotherm-psd}b). The desorption branches differ between the hardened  and the two relaxed configurations, showing a densification tendency altering pore constrictions, due to structural relaxation. We note that the distribution of pore sizes between the hardened and relaxed configurations show no significant differences (Fig.\ref{fig:sorption-isotherm-psd}d).

Pore size distribution (PSD) is an important piece of information in the characterization of porous media, which experimentally is often derived from adsorption/desorption isotherms \citep{lowell2012characterization, sing1985reporting, rouquerol2012characterization}, using Kelvin's equation and surface adsorption models\citep{barrett1951determination, hagymassy1969pore, langmuir1918adsorption}. However, isolated and geometrically idealized pore shapes such as cylinders or spheres are usually assumed in these calculations. Another way to calculate PSD is using Monte Carlo (MC) sampling of the pore space of simulated samples (as in this work) or experimentally-acquired 3d tomography images \citep{bhattacharya2006fast}. To access the pore connectivity down to nanometer scale, techniques as nuclear magnetic resonance (NMR) or X-ray tomography (see e.g. \citep{rouquerol2012characterization, chae2013advanced}) are used. Here, we apply the MC algorithm to the mesoscale cement paste models that have a realistic PSD in agreement with experimental data \cite{ioannidou2016mesoscale}.

The extracted PSDs are compared in Fig.\ref{fig:sorption-isotherm-psd}c and d for nitrogen and water respectively. The MC results capture a population of gel pores ($\sim2$~nm) and large pores that extends to $\sim$ 25~nm. The Kelvin equation results for nitrogen shows similar features. However, for water the Kelvin equation results feature bi-modal distributions with both peaks smaller than 5~nm, similar to the curves derived from experimental data in Ref. \citep{baroghel2007water}. The lack of large pores derived from the Kelvin equation using water sorption isotherm data is due to the non-linear relationship between Kelvin radius and the relative humidity $RH$, so that pores larger than 5~nm will correspond to $RH$>95\%. Our results indicate that the PSD derived from nitrogen soprtion isotherms are consistent with the real morphology of the pore space, while those derived from water sorption isotherms lack the large capillary pores. However, such incomplete PSDs from water sorption experiments are often reported in the literature.

\subsection*{Capillary Forces at nano-grains Level}

Having the water distributions from the lattice gas DFT simulations, we proceed to calculate the capillary forces on the nano-grains following our previous work described in Ref.~\change{\citep{capillary-langmuir}. At low water saturation, capillary bridges between 2 or more nano-grains can be described by Kelvin's equation. This traditional method requires increasing effort for more nano-grains~\citep{melnikov2015grain, delenne2015liquid} and fail when liquid clusters starts percolating. The advantage of our method~\citep{capillary-langmuir} is to enable us to calculate capillary stress at arbitrary liquid saturation levels.} We first construct the Korteweg stress field \citep{korteweg1901forme,anderson1998diffuse} based on the liquid density $\rho$ and its gradient $\grad\rho$ \change{(see Materials and Methods).}
Then we integrate the stresses over the surface of Voronoi cells of each nano-grains to obtain the capillary forces.

\begin{figure}[!t]
	\centering
	\begin{subfigure}{0.95\linewidth}
		\centering
		\includegraphics[width=1.0\linewidth]{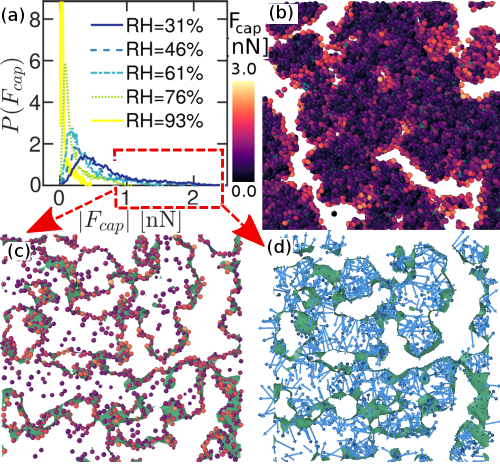}
	\end{subfigure}
	\caption{
		\label{fig:fcap-on-particles}
		Distribution and visualization of capillary forces.
		(a) Distribution of capillary forces $\vert F_{cap} \vert$ at the particle level. (b) Color code shows the magnitude of $F_{cap}$ on a thin section of the sample at $RH=31\%$. The large forces are concentrated on the surface of pores, as highlighted in (c) (d). The green segments in (c) and (d) compose the pore-solid interface. In (c) only particles having $F_{cap}>1.0$~nN are shown in the same color code as (b), while (d) is showing only force vectors in light blue. Most of these forces are pointing into the solid packing region, effectively densifying the gel texture. The simulation box size is 400~nm.}
\end{figure}

Fig.\ref{fig:fcap-on-particles}a shows the water capillary force \change{per nano-grain} distributions computed at different $RH$ values. Qualitatively, capillary forces increase in magnitude when $RH$ decreases. The forces at $RH=31\%$ (the closing point of the hysteris loop) exhibit a long-tailed distribution at the particle level,  with a mean value around $0.7$~nN and the largest $5~\%$ is above $1.8$~nN (see Fig.\ref{fig:fcap-on-particles}a). The maximum of $\vert F_{cap}\vert$ (3~nN) is still smaller than the magnitude of solid-solid particle interaction forces ($ F_{grain} \approx$ 5-10~nN). Fig.\ref{fig:fcap-on-particles}b shows the magnitude of the capillary force on each particle of a thin section of the 3D hardened cement paste.

Fig.\ref{fig:fcap-on-particles}c and d depict the particles experiencing the largest capillary forces at $RH=31\%$ (magnitude and vector respectively). \change{The small gel pores are always filled until low $RH$, while large capillary pores remain empty even at high $RH$, thus creating largest capillary forces} between the gel pores (1$\sim$\change{5}~nm) and the capillary pores (>\change{5}~nm). The overall capillary effect at $RH=31\%$ is densification of the solid texture since the capillary forces enhance attractions between nano-grains surrounding small capillary pores.

\subsection*{\change{Irreversible} Rearrangements due to Capillary Stresses}

\begin{figure}[!t]
	\centering
	\includegraphics[width=1.0\linewidth]{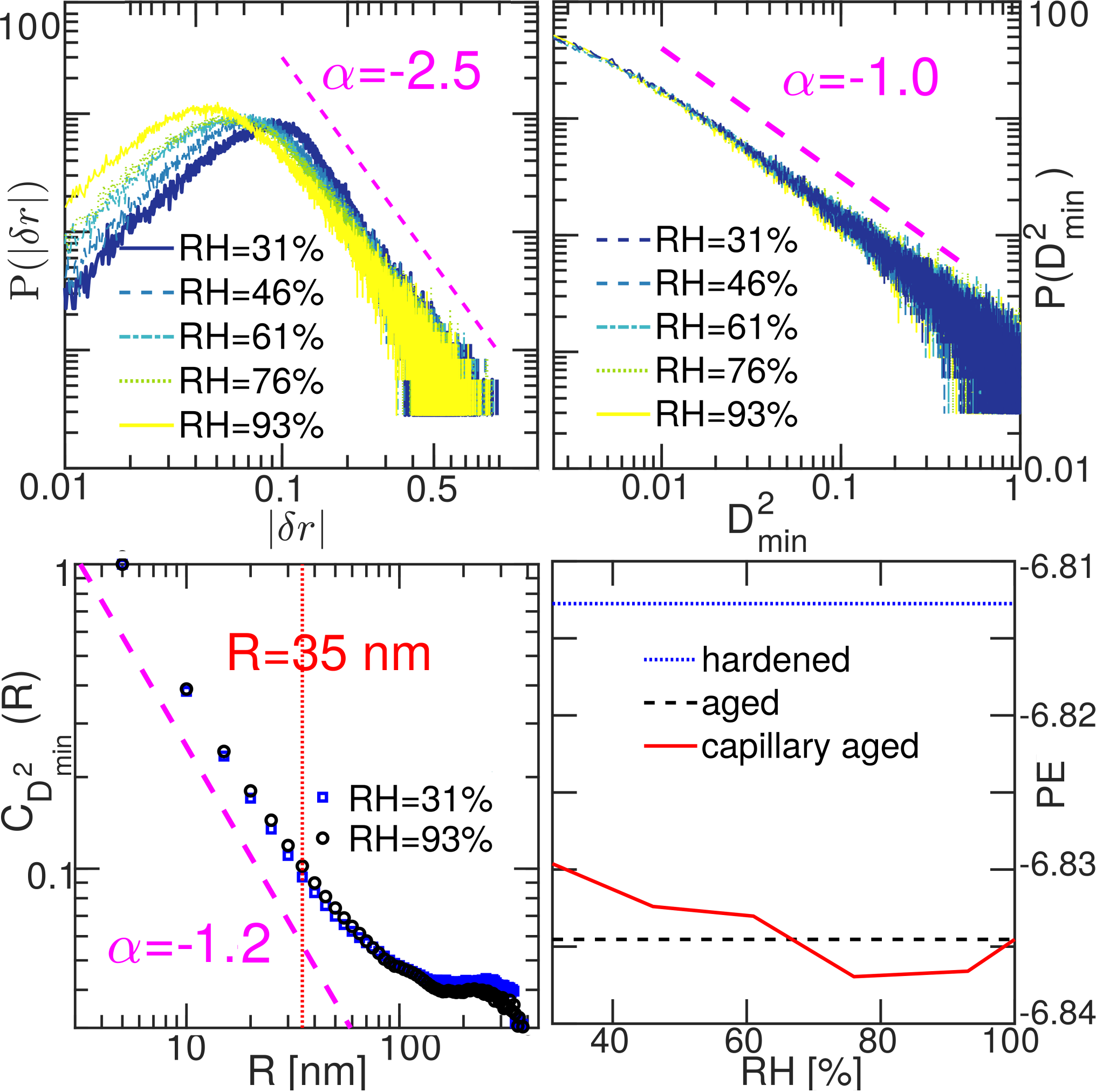}
	\caption{
		\label{fig:plasticity}
		Structural relaxation and \change{nonaffine displacements} due to capillary forces.
		(a) Distributions of particle displacement magnitudes $\vert \delta r\vert=\vert \Delta \vec{r} / \sigma \vert$ ($\sigma$ is average particles diameter) at different $RH$'s.
		(b) Distributions of nano-grains nonaffine displacements $D_{min}^2$. The aged sample is the reference state.
		(c) Spatial correlation function of $D_{min}^2$. Power-law fits are shown in dashed lines.
		(d) Potential energy (PE) after structural relaxation simulated with and without capillary forces (aged and capillary aged respectively) at different $RH$ values. The initial PE of the hardened paste before relaxation is also shown. For $RH>70\%$ capillary stress assists the relaxation process and lower the PE of the system. PE values are in reduced units.}
\end{figure}

We further investigate the relaxation of the solid structure with eigenstresses (hardened sample) under wet conditions using MD simulations with capillary and pairwise particle interactions. We assume the capillary forces to be constant along the MD trajectories. This assumption is proven to be valid as capillary forces do not induce large particle displacements as shown in Fig.\ref{fig:plasticity}(a) where the distributions of particle displacement \change{magnitudes ($\delta r=\vert\Delta \vec{r}/\sigma\vert$, $\sigma$ being the average particle diameter)} are displayed. Under the action of capillary forces in all $RH$, only <0.1\% of the particle population undergo displacement comparable to their own size (i.e. few nm). This is due to the fact that the magnitude of the capillary forces are in general five times smaller than inter-particle cohesion. As the $RH$ value decreases, capillary forces increase, and the peak of $\delta r$ distribution shifts to larger values. At larger displacements, the distributions at all $RH$ exhibit an algebraic decay with characteristic exponent equal to $-2.5$.

To investigate further the effect of capillarity in the structural relaxation of cement, we measure the \change{drying induced irreversible} diplacements in the capillary aged sample taking as reference the aged sample. \change{This quantity is akin to what in glass physics is coined nonaffine displacement, which in our case is a result of capillary pressure. (See more details in Materials and Methods.)}

Fig.{\ref{fig:plasticity}}(b) shows the distributions of \change{nonaffine deformations} $D_{min}^2$ that exhibit long range correlations with algebraic decay with exponent of $-1$ at all $RH$.
However, further analysis of $D_{min}^2$ by means of the spatial correlation function
\begin{equation}
C_{D_{min}^2}(R)=\frac{\left\langle D_{min}^2(r)D_{min}^2(r+R) \right\rangle - \left\langle D_{min}^2(r) \right\rangle^2}{\left\langle D_{min}^2(r)^2 \right\rangle - \left\langle D_{min}^2(r) \right\rangle^2}
\end{equation}
shows an algebraic behavior of exponent $\alpha\sim-1.2\pm 0.1$ restricted to short distances (R<35~nm, see Fig.\ref{fig:plasticity}(c)) by contrast to observations for dense colloidal glasses under shear deformations \citep{chikkadi2012nonaffine, chikkadi2011long}. We tentatively attribute this behavior to the presence of capillary pores that localized these \change{nonaffine deformations} at their vicinity.

We note that high RH values (>$70\%$) result in a lower potential energy (PE) of the system than the fully wet aging condition (no capillary effects) after relaxation, as seen in Fig.3(d). This range of $RH$ values corresponds to capillary forces of smaller amplitude  (see Fig.2a). Capillary forces of larger amplitude at RH <$70\%$ increase the PE.

\subsection*{Stress Homogenization}
\label{sec:stress-homogenization}

\begin{figure*}[!t]
	\centering
	\includegraphics[width=1.0\linewidth]{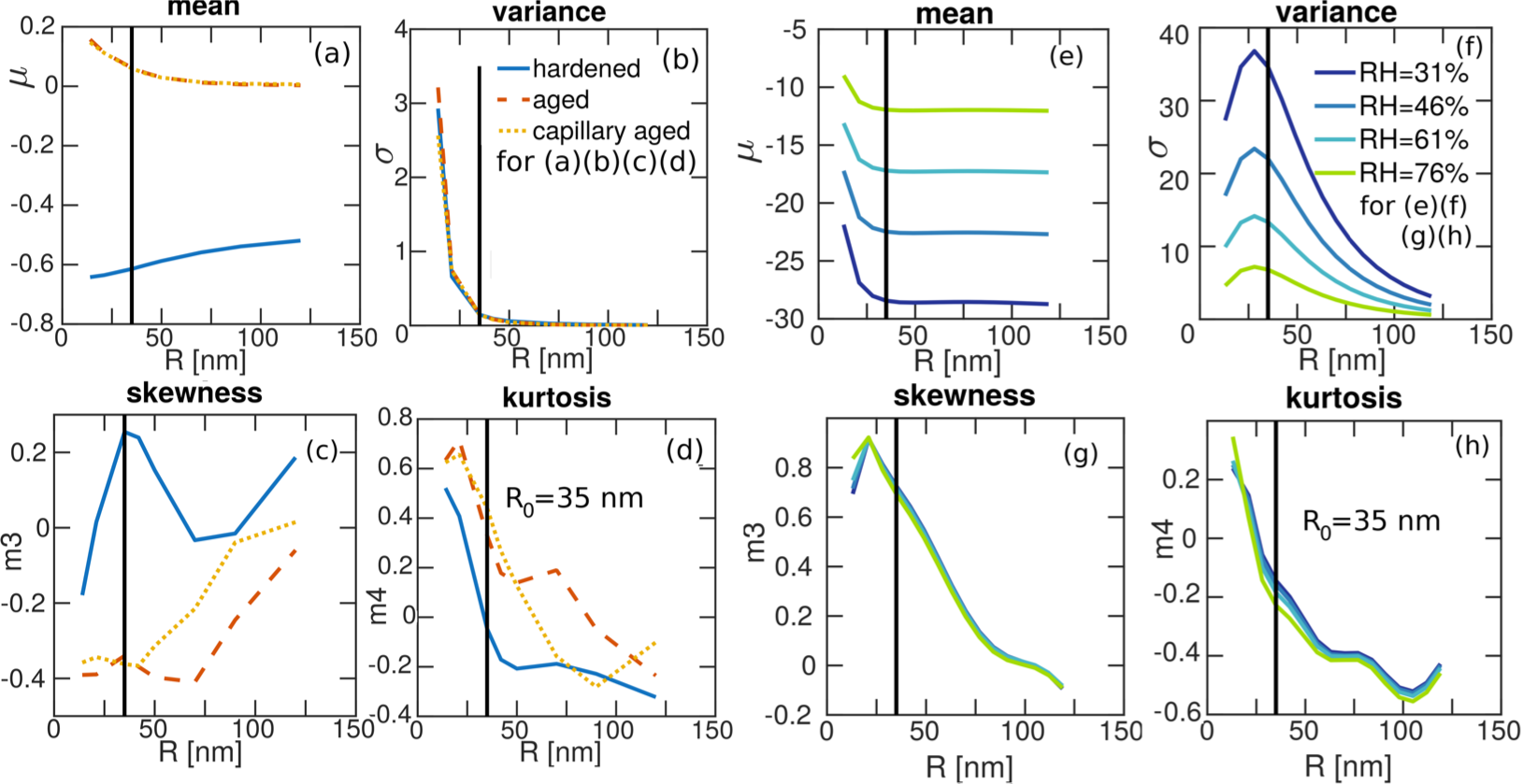}
	\caption{
		\label{fig:homo-scalings}
		Homogenization length scales for solid and liquid stresses.
		(a-d) show the statistics of solid volumetric stress ($P_{\mathrm{vol}}$) in hardened, aged and capillary aged samples. (e-f) show the statistics of liquid pressure $P_{\mathrm{vol}}$.
 Black lines denote the homogenization length R$_0$=35~nm, common for all samples/stresses. The simulation boxes are $\sim$ (400~nm)$^3$.
	}
\end{figure*}

The capillary forces calculated on the cement hydrate nano-grains show a heterogeneous distribution (Fig.\ref{fig:fcap-on-particles}(a)).
We adopt the Love-Weber homogenization\citep{love2013treatise, weber1966recherches} scheme to coarse-grain the nanometric capillary stress distribution (see SI Appendix).
For solid and capillary stresses, we tested different homogenization scales (Fig.\ref{fig:homo-scalings}) ranging from 10~nm to 100~nm in all the cement paste models considered in this work. The distribution of solid stress in the hardened sample shows a maximum at non zero pressure due to the eigenstress acquired during setting. The distributions of solid stresses in the aged and capillary aged samples are similar in amplitude and width (see SI Appendix). The shape and width of the capillary stress distribution at $RH=31\%$ for the hardened sample change with $R$ (see Fig.S3 and skewness in Fig.\ref{fig:homo-scalings}g ).  However, Fig.\ref{fig:homo-scalings} presenting the first four moments of these distributions, indicates that  a characteristic  length scale emerges around $R_0\sim$35~nm. Above this scale, the first moments converge to the macroscopic mean, the second moments monotonically decreases and the excess kurtosis is negative or close to zero indicating shorter tails than the normal distribution.

Beside the heterogeneity of the structure, it is interesting to note that $R_0$ is an order of magnitude smaller than the simulation box size ($\sim 400$~nm) and it is also similar to the correlation length of $D_{min}^2$ in Fig.\ref{fig:plasticity}(c). The emerging homogenization length common to both capillary stress and solid stress, is rationalized as the scale of long range spatial correlations inherent in the heterogeneous texture has been shown in Ref. \citep{ioannidou2016mesoscale} via scattering data and chord length distributions.

\begin{figure*}[!t]
	\centering
	\includegraphics[width=1.0\linewidth]{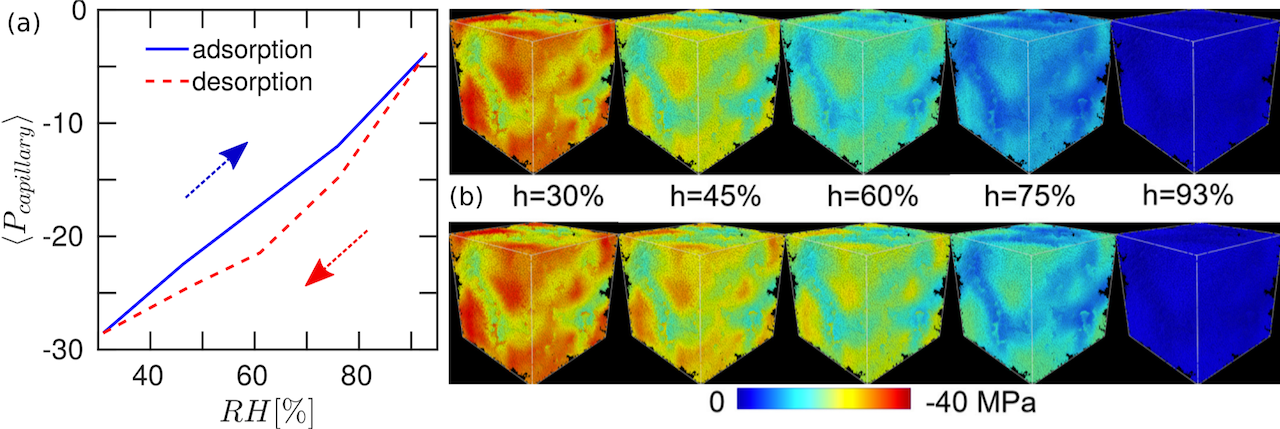}
	\caption{
		\label{fig:pcap-hysteresis}
		Homogenized stresses.
		(a) Hysteresis of average capillary pressure. Negative sign indicates shrinkage.
		(b) Hysteresis of homogenized capillary pressure displayed at the meso-scale. Upper row for adsorption, lower row desorption. The simulation boxes are $\sim $(400~nm)$^3$. Homogenization scale $R_0\sim $35~nm.
	}
\end{figure*}

Averaging over the entire simulation box results in the overall capillary pressure  at the micron level, as shown in Fig.\ref{fig:pcap-hysteresis}(a).
\change{Cement paste, although a heterogeneous porous medium, displays a hysteretic capillary pressure}
similar to that found for simple geometry porous media \citep{gor2010adsorption, gor2017adsorption}. Fig.\ref{fig:pcap-hysteresis}(b) shows the homogenized capillary pressure calculated at $R_0$=35~nm at different $RH$ during adsorption (upper row) and desorption (lower row). In absolute value, the homogenized capillary stress is larger upon desorption and it is more intense in densest matrix regions.

\subsection*{Conclusion}
In this work, we investigated the role of capillary forces in the relaxation mechanism of a granular model of cement paste under partial saturation, from the nano-grain level (5~nm) to the mesoscale (400~nm).
\change{We achieved agreement with nitrogen and water sorption experiments, and assessed the validity of the Kelvin equation in determining the PSD of cement paste. We showed that nitrogen desorption data give better PSD than water, compared with the PSD obtained from 3D stereological analysis. We also considered one more drying/wetting cycle showing structural densification but without significant change of the PSDs.}

\change{
The framework that we developed based on the liquid distribution and stress field at the nano-grain level enables us to show that the largest capillary forces are concentrated at the interface between gel pores and larger capillary pores.
We found that the largest capillary forces are obtained at the closure point of the hysteresis loop (RH=31\%) and correspond to menisci located at the interface between gel and capillary pores.
These forces induce local \change{irreversible} rearrangements that
have relatively short spatial length correlations due to the presence of the capillary pores spanning the samples, and hence do not exhibit the long-range character found in colloidal glasses during inhomogeneous shear \citep{chikkadi2011long, chikkadi2012nonaffine}. Under additional mechanical loading and longer timescales, \change{or with multiple drying/wetting cycles}, we speculate that they may contribute altogether to creep and fracture, similar to observations in amorphous materials~\citep{cao2017understanding, cao2018nanomechanics} and fluid saturated rocks\citep{brantut2011pore}.
}

\change{
The results suggest a homogenization length scale common to both liquid and solid stress fields, correlated with the scale of structural heterogeneities. Therefore this length sets the representative volume element (RVE) size for upscaling poromechanics, either using discrete or continuum descriptions such as the LEM method\cite{laubie2017stress}.
}

In soft porous materials, capillary forces are comparable to the solid cohesive forces and may induce large deformations, such as drying of colloidal suspensions~\citep{dufresne2003flow, tsapis2005onset}. The framework we proposed can be readily adopted in these materials.

\matmethods{

\subsection*{Sorption Simulations and Stress Calculation}
The sorption simulations are  based on the lattice gas density functional theory model~\citep{kierlik2001capillary, kierlik2002adsorption}, by minimizing the grand potential:
$
\Omega = -w_{ff}\sum_{<i,j>}\rho_i\rho_j -
yw_{ff}\sum_{i,j}\rho_i\eta_j -
\mu\sum_i \rho_i
+ k_BT \sum_i \left[\rho_i \ln\rho_i +
(1-\rho_i)\ln(1-\rho_i)\right]
$,
where $\rho_i\in [0,1]$ denotes the normalized density of fluid on site i, continuously varying from 0 to 1, $\eta_i=0$ or 1 indicating site i occupied by solid or vacant, $w_{ff}$ and $w_{sf}$ are the fluid-fluid interaction and fluid-solid interaction energy, respectively. $w_{ff}$ is determined by the bulk critical point $k_B T_c = -\nu w_{ff}/2$, with $\nu$ the number of nearest neighbors. $w_{sf}$ is imported from atomistic simulation data as the isosteric heat of adsorption (i.e. the differential heat of adsorption), in the limit of zero-coverage.
For water and nitrogen adsorption in cement $w_{ff}/w_{sf}=2.5$ \cite{bonnaud2012thermodynamics}. To estimate the lattice spacing $a$, we estimate the surface tension which is energy per area, $\gamma\sim w_{ff}/2a^2$. For nitrogen at $T=77$~K, $\gamma\sim$8.94~mN/m which gives $a\sim$0.345~nm. For water at $T=300$~K, $\gamma\sim$72~mN/m which gives $a\sim$0.24~nm. Based on the estimates we choose a fine grid cell size of $a=3\AA$. Taking $\rho=2.5$~g/mL for solid phase density of cement hydrates, we convert degree of saturation measured from the simulations to total amount of absorbed liquid in cement paste.
\change{The Korteweg stress tensor can be derived as (see SI Appendix)
$\tensor{\bm\sigma} =  \left( p_0(\rho) - \frac{a^2 w_{ff}}{2} (\vec{\nabla}\rho)^2 \right) \tensor{\bm I} + a^2 w_{ff} \vec{\nabla}\rho \otimes \vec{\nabla}\rho + \tensor{\bm\sigma}_0
$,
where
$\tensor{\bm I}$ is the identity tensor and $\tensor{\bm\sigma}_0$ an arbitrary constant tensor. $p_0(\rho)=\mu\rho + \frac{\nu w_{ff}}{2}\rho^2 - k_BT \left[ \rho \ln (\rho) + (1-\rho) \ln (1-\rho) \right]$ is the asymptotic bulk value of the hydrostatic pressure.}

\subsection*{Simulations of structural relaxation}
The initial mesoscale configuration of cement hydrates is the hardened configuration that has an average eigenstress $\sim$ -50~MPa. Structural relaxation MD simulations of the hardened sample were carried out using LAMMPS\citep{plimpton1995fast} on NPT conditions of zero pressure and room temperature. Capillary forces at different $RH$ values were constantly applied during the entire simulation. The hardened configuration under the action of capillary forces at $RH=31\%$, where the distribution of capillary forces has the longest tail is called capillary aged. All simulations were terminated after 500000 MD steps with time step $\delta t=0.0025$ in reduced unit when the system had converged. From the particle positions, nonaffine displacements were computed.
\change{The nonaffine displacement is defined as $D^2_{min} = \min_{\tensor{F}}{\sum_i \left[\vec{r}_i - \vec{r}_0 - \tensor{F} \cdot (\vec{r}_i^{ref} - \vec{r}_0^{ref})\right]^2}$, where $i$ runs over neighbors of the central particle indexed by 0~\citep{falk1998dynamics, shimizu2007theory}. $\tensor{F}$ corresponds to the deformation gradient in continuum mechanics, and $D_{min}^2$ essentially captures the higher order deformations that are usually not considered in linear theories.}
The reference configuration for the nonaffine displacements was the aged configuration. The aged configuration is final state of the hardened configuration after relaxation to zero average eigenstress by MD simulation in the same NPT conditions (zero pressure and room temperature) but without capillary forces (only inter particle interactions).

}

\showmatmethods

\section{Supplement Information Appendix}

\subsection{Mesoscale model of hardened cement paste}
Ioannidou's \textit{et al.} mesoscale C-S-H model \cite{ioannidou2017inhomogeneity,ioannidou2016mesoscale} was used to calculate the water adsorption/desorption by DFT simulations. The precipitation of C-S-H nano-grains and settings was simulated using the approach recently proposed in Ref. \cite{ioannidou2014controlling}. In this approach, a free energy gain drives the precipitation of particles of few nanometers (C-S-H nanoscale hydrates) which also interact and aggregate. The simulations consisted of a Grand-Canonical Monte Carlo (GCMC) scheme, where the chemical potential corresponds to the free energy gain just mentioned, coupled to a Molecular Dynamics (MD) scheme.

The effective interparticle forces between cement hydrates depends on the concentration of calcium ions in the solution and changes during the hydration \cite{roland2004does,Plassard-lang2005}. In precious works, the microstructure of C-S-H gels at early hydration stages was investigated using attracto-repulsive potential arising from ion-ion correlation forces \citep{ioannidou2014controlling, ioannidou2017inhomogeneity}. 

Differently from \cite{ioannidou2014controlling} here we consider only effective interactions that would correspond to the hardened cement paste and can be well modeled \cite{ioannidou2016mesoscale} with a short-range attractive Mie potential 
\begin{equation}
V(r)=\alpha \epsilon \left[ \left( \frac{\sigma}{r}\right)^{2\gamma} - \left( \frac{\sigma}{r}\right)^{\gamma}  \right] \ , \label{eq:pot}
\end{equation}
where $r$ is the inter-particle distance, $\alpha$ is the well depth (with $\epsilon$ the unit energy) between two particles with size diameter $\sigma$ and we have fixed the exponent to $\gamma=12$. We have set $\alpha=6$ and the temperature to $T=0.15$ (typically measured in units $\epsilon/k_{B}T$), while time is measured in usual MD units $\sqrt{m\sigma^2/\epsilon}$, all in reduced units.

A GCMC cycle consists of $N_{MC}$ attempts of particle insertion or deletion and it is followed by $N_{MD}=100$ MD steps. $R=N_{MC}/(N_{MD} \cdot L^3 \cdot \delta t)$ where $L=390.36$ nm is the length of the simulation box and $\delta t =0.0025 \sqrt{m \sigma^2/\epsilon}$, is the rate of hydrate production that mimics the chemistry of the system. For the simulations reported here we use a chemical potential $\mu=-1$ (in reduced units) and a rate $R=25 \cdot 10^{-9} \delta t^{-1} nm^{-3}$. 

During hydration of cement, densification of C--S--H is heterogeneous and that this is the source of significant local structural and mechanical heterogeneitieis in the final solid material. For a more realistic structure, particle size polydispersity was introduced in the following way. Each particle of a configuration containing 177975 particles was randomly assigned a particle diameter in the range between $3.78$ and $9.2$ nm. To facilitate the computation the range of diameters was discretized into 53 intervals. Then the simulation box size was increased by $(\sigma_{max}+\sigma_{min})/2$ to avoid particle overlapping and energy minimization with the conjugate gradient algorithm was applied. The mesoscale structure produced by the precipitation persists after the introduction of polydisperse particle sizes. All MD and GCMC simulations have been performed using LAMMPS \cite{plimpton1995fast}.

The model cement paste has accumulated eigen-stresses due to the out-of-equilibrium precipitation of nano-grains. This configuration is named ``hardened'' cement paste. This particle configurations was relaxed to a total pressure as close to zero as possible, being less than $2 \cdot 10^{-6}$ $\epsilon/\sigma^3$. Cycles of energy minimization using conjugate gradient with and without box size change were applied with energy tolerance $10^{-10}$. This relaxed configuration is named ``aged'' cement paste.
The hardened sample relaxed with capillary forces
calculated on the hardened structure at RH = 31\%
where capillary effects are larger, referred to as “capillary
aged” sample.

\subsection{Sorption simulation and capillary forces}

We use the lattice gas density functional theory (DFT) to simulate both nitrogen and water adsorption/desorption isotherms in cement paste. This approach was first developed by Kierlik $et~al$ \citep{kierlik2001capillary, kierlik2002adsorption} for adsorption/desorption of a non polar fluid in a quenched random porous solid. Subsequently, it was applied to Vycor\citep{coasne2013adsorption}, controlled porous silica glasses and aerogels \citep{detcheverry2004mechanisms, kierlik2001equilibrium}, to infer qualitatively adsorption/desorption isotherms through minimizing the grand potential:
\begin{equation}
\begin{split}
\Omega = & -w_{ff}\sum_{<i,j>}\rho_i\rho_j -
yw_{ff}\sum_{i,j}\rho_i\eta_j -
\mu\sum_i \rho_i \\
& + k_BT \sum_i \left[\rho_i \ln\rho_i +
(1-\rho_i)\ln(1-\rho_i)\right]
\end{split}
\label{main eqn:grand potenial}
\end{equation}
where $\rho_i\in [0,1]$ denotes the normalized density of fluid on site i, continuously varying from 0 to 1, $\eta_i=0$ or 1 indicating site i occupied by solid or vacant. $w_{ff}$ and $w_{sf}$ are the fluid-fluid interaction and fluid-solid interaction, respectively. $w_{ff}$ is only a property of the fluid, determined by the bulk critical point $k_B T_c = -n w_{ff}/2$, with $n$ the number of nearest neighbors. While $w_{sf}$ is imported from atomistic simulation data as the isosteric heat of adsorption (i.e. the differential heat of adsorption), in the limit of zero-coverage (for water adsorption in cement see Figure 4 in \cite{bonnaud2012thermodynamics}, giving $w_{ff}/w_{sf}=2.5$. Same ratio is taken for nitrogen).
The parameters $w_{ff}$ and $w_{sf}$ are in units of energy. To arrive at a characteristic length scale as the lattice spacing a, we estimate the surface tension which is energy per area:
\begin{equation}
\gamma\sim \frac{w_{ff}}{2a^2}
\end{equation}
for nitrogen at $T=77~K$, $\gamma\sim8.94mN/m$ which gives $a\sim0.345~nm$; for water at $T=300~K$, $\gamma\sim72mN/m$ which gives $a\sim0.24~nm$. Based on the estimates we choose a fine grid cell size of $a=3\AA$.
Taking $\rho=2.5~g/mL$ for solid phase density of CSH, we convert degree of saturation measured from the simulations to total amount of absorbed water in cement paste.

For capillary force and drying shrinkage, we first simulate one wetting-drying cycle of water sorption on the hardened structure with eigen-stress that has not been released after C-S-H precipitation. Then we take liquid distributions at different relative humidities and calculate capillary forces on the nano-particles according to the framework described in reference \citep{2018arXiv180305879Zhou}. The continuum limit of the lattice gas model is the Cahn-Hilliard model:
\begin{equation}
\begin{split}
g({\rho}) & = \int\limits_V dV \left\lbrace -\frac{n}{2} w_{ff}\rho^2 + \frac{a^2}{2} w_{ff}(\grad\rho)^2 - \mu\rho \right\rbrace \\
& + \int\limits_V dV \left\lbrace k_BT \left[\rho \ln\rho +
(1-\rho)\ln(1-\rho)\right] \right\rbrace  \\
& + \int\limits_{\partial V} d\vec{S}\cdot \left( w_{sf}\rho \vec{n}_{sf} - \frac{a^2}{2} w_{ff} \rho\grad\rho \right)
\end{split}
\end{equation}
where $\vec{n}_{sf}$ is the boundary normal vector at liquid-solid surface denoted by $\partial V$, pointing from liquid outward into solid; $a=0.3~nm$ is the choice of lattice spacing, $n$=6 the number of nearest neighbors. We can derive the Korteweg stress tensor in this context. Thermodynamics shows that pressure is $p=\mu\rho-f_0(\rho)=\rho f_0'-f_0=\frac{1}{3}\Tr\tensor{\sigma}$ for a homogeneous system of free energy $f_0$. 
At mechanical equilibrium, we generalize the stress tensor for an inhomogeneous system, insisting that it should satisfy:
\begin{equation}
-\grad \cdot(\tensorsigmanod-\mu\rho\tensoriden) = \grad (\rho f_0') - \rho\grad f_0' = \grad\cdot (f_0\tensoriden) = \frac{\delta f_0}{\delta \rho} \grad \rho
\end{equation}
where the last equality essentially neglects all higher order terms in the expansion of $f_0$.
Using Stokes' theorem
\begin{equation}
\begin{split}
\int f_1 = & \int \left(\grad g_1(\rho)\right)^2 \\
= & \int \grad\cdot\left(g_1(\rho)\grad g_1(\rho)\right)-g_1(\rho)\grad^2g_1(\rho)\\
= & -\int g_1(\rho)\grad^2g_1(\rho)
\end{split}
\end{equation}
Now $f(\rho, \grad \rho)=f_0+f_1^*$ where $f_1^*=-g_1(\rho)\grad^2g_1(\rho)$
Because of the identities
\begin{equation}
\begin{split}
\vec{u}\times(\grad \times\vec{u}) = & \frac{1}{2}\grad (\vec{u}\cdot\vec{u}) - \vec{u}\cdot\grad \vec{u}\\
\vec{u}\cdot\grad \vec{u} = & \grad \cdot(\vec{u}\otimes\vec{u}) - \vec{u}\grad \cdot\vec{u} \\
\grad \left(\vec{u}\cdot\vec{u}\right) = & \grad \cdot \left( \vec{u}\cdot\vec{u} \tensoriden \right)
\end{split}
\end{equation}
and notice $\grad\times\grad g_1(\rho)=0$ we then have
\begin{equation}
\begin{split}
\frac{\delta f_1^*}{\delta \rho}\grad\rho & = -2\grad g_1(\rho)\grad^2g_1(\rho)
\\
= & -2 \left[ \grad\cdot\left( \grad g_1(\rho)\otimes\grad g_1(\rho) \right) - \grad g_1(\rho)\cdot\grad \grad g_1(\rho) \right] 
\\
= & -2 \grad\cdot \left[ \grad g_1(\rho)\otimes\grad g_1(\rho)  + \grad g_1(\rho)\cdot\grad g_1(\rho) \tensoriden \right]
\end{split}
\label{eqn-stress:core-derivative}
\end{equation}
Finally we arrive at the generalized inhomogeneous stress tensor
\begin{equation}
\begin{split}
\tensorsigma = & 2 \grad g_1(\rho)\otimes\grad g_1(\rho) + \left( \mu\rho - (\grad g_1(\rho))^2 \right)\tensoriden + \tensorsigma_0 \\
= & \left( p_0(\rho) - \frac{a^2 w_{ff}}{2} (\vec{\nabla}\rho)^2 \right) \tensor{\bm I} + a^2 w_{ff} \vec{\nabla}\rho \otimes \vec{\nabla}\rho + \tensorsigma_0
\end{split}
\end{equation}
where $\tensorsigma_0$ is an arbitrary tensor constant and $p_0(\rho)=\mu\rho + \frac{\nu w_{ff}}{2}\rho^2 - k_BT \left[ \rho \ln (\rho) + (1-\rho) \ln (1-\rho) \right]$ the asymptotic bulk value of the hydrostatic pressure.
Integration of the stress tensor over Voronoi cell faces of nano-grains produces a capillary force vector associated with each nano-grain.

\subsection{Relaxation with and without capillary forces}

For each relative humidity ($RH$) value, molecular dynamics (MD) simulations were carried out using LAMMPS\citep{plimpton1995fast, plimpton2007lammps} on the cement hydrate model, performing NPT relaxation at room temperature and 0 ambient pressure. NVT relaxation are also performed but not discussed here. Capillary forces were constantly applied during the entire simulation. All simulations were terminated after 500000 MD steps with timestep $\delta t=0.0025$ in Lennard-Jones unit when the system is already converged and stable.
We start from the configuration right
after out-of-equilibrium dissolution-precipitation, referred to as “hardened” sample, that has an overall eigenstress $\sim$ -50~MPa. 2) the hardened sample relaxed
to 0 average eigen-stress, referred to as “aged” sample
and 3) the hardened sample relaxed with capillary forces
calculated on the hardened structure at RH = 31\%.

\begin{figure}
\centering
\includegraphics[width=0.94\linewidth]{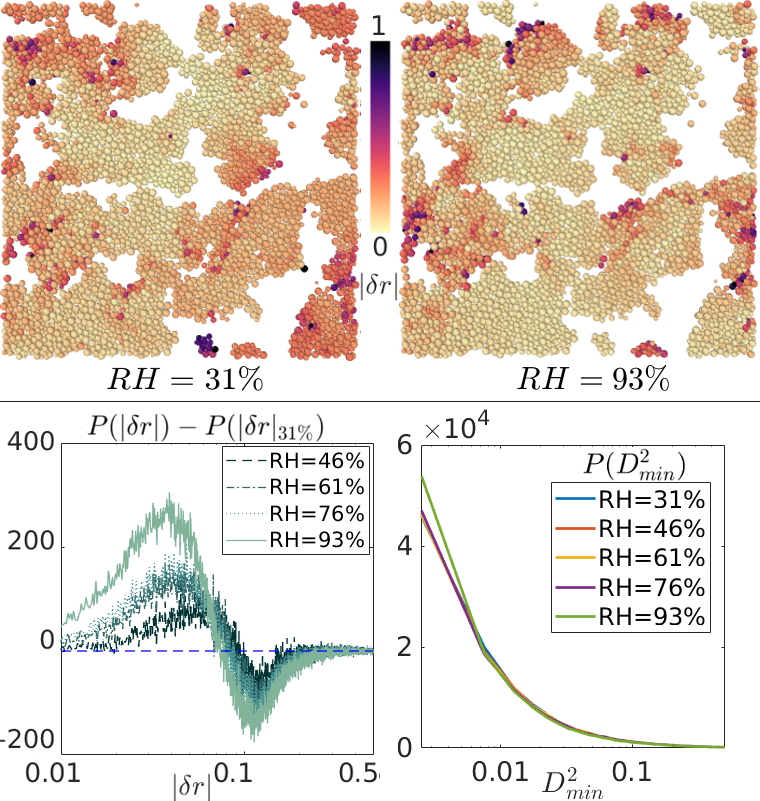}
\caption{
\label{fig:SI-displacementmag}
Upper panels show a cross section colored by displacement magnitudes $\vert \delta r\vert$. About 0.1\% nano-particles (148 out of 177975) are displaced $\vert\delta r\vert>1$. 
Lower left panel shows the difference between histograms at higher $RH$ values and the one at $RH=31\%$.
Lower right panel shows the histograms of $D_{min}^2$ (see text for definition) at various relative humidities and the differences with respect to the one at $RH=31\%$.
Cut-off radius is chosen as 2 times of average nano-particle radii, about $3\sim4\%$ times of the nano-particles having $D_{min}^2>1$.
}
\end{figure}

\begin{figure}
\centering
\includegraphics[width=0.94\linewidth]{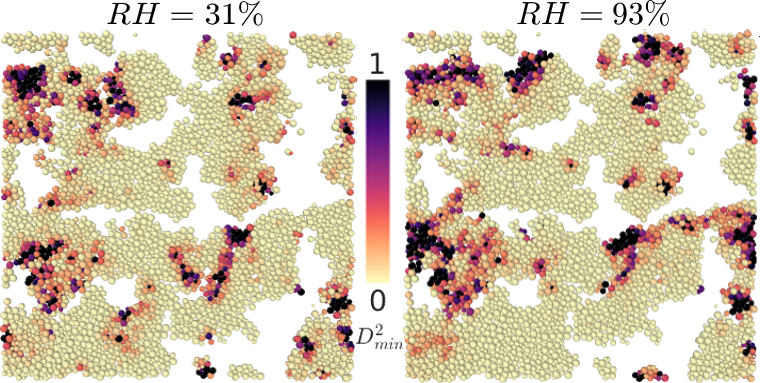}
\caption{
\label{fig:SI-Dmin2-crosssection}
$D_{min}^2$ (see text for definition) at various relative humidities shown on a cross section of the mesoscale cement model.
Cut-off radius is chosen as 2 times of average nano-particle radii.
}
\end{figure}

\begin{figure}
\centering
\includegraphics[width=0.94\linewidth]{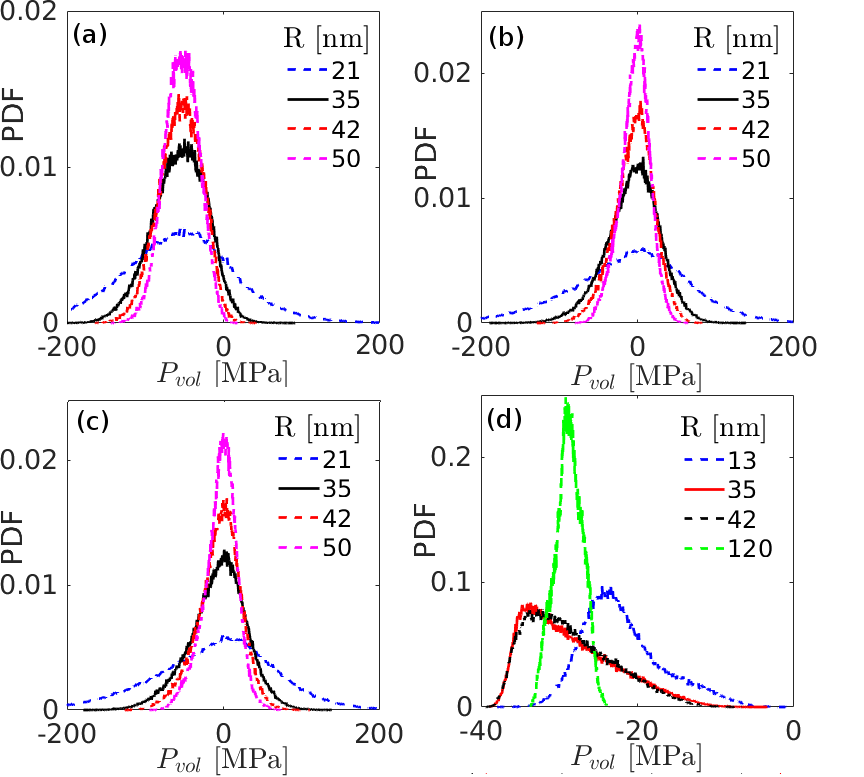}
\caption{
\label{fig:SI-homo-pdf}
The distributions of homogenized stresses for (a) solid volumetric stress in hardened sample, (b) solid volumetric stress in aged sample, (c) solid volumetric stress in capillary aged sample, and (d) liquid volumetric stress  at RH=31\%.
}
\end{figure}

\begin{figure}
\centering
\includegraphics[width=0.94\linewidth]{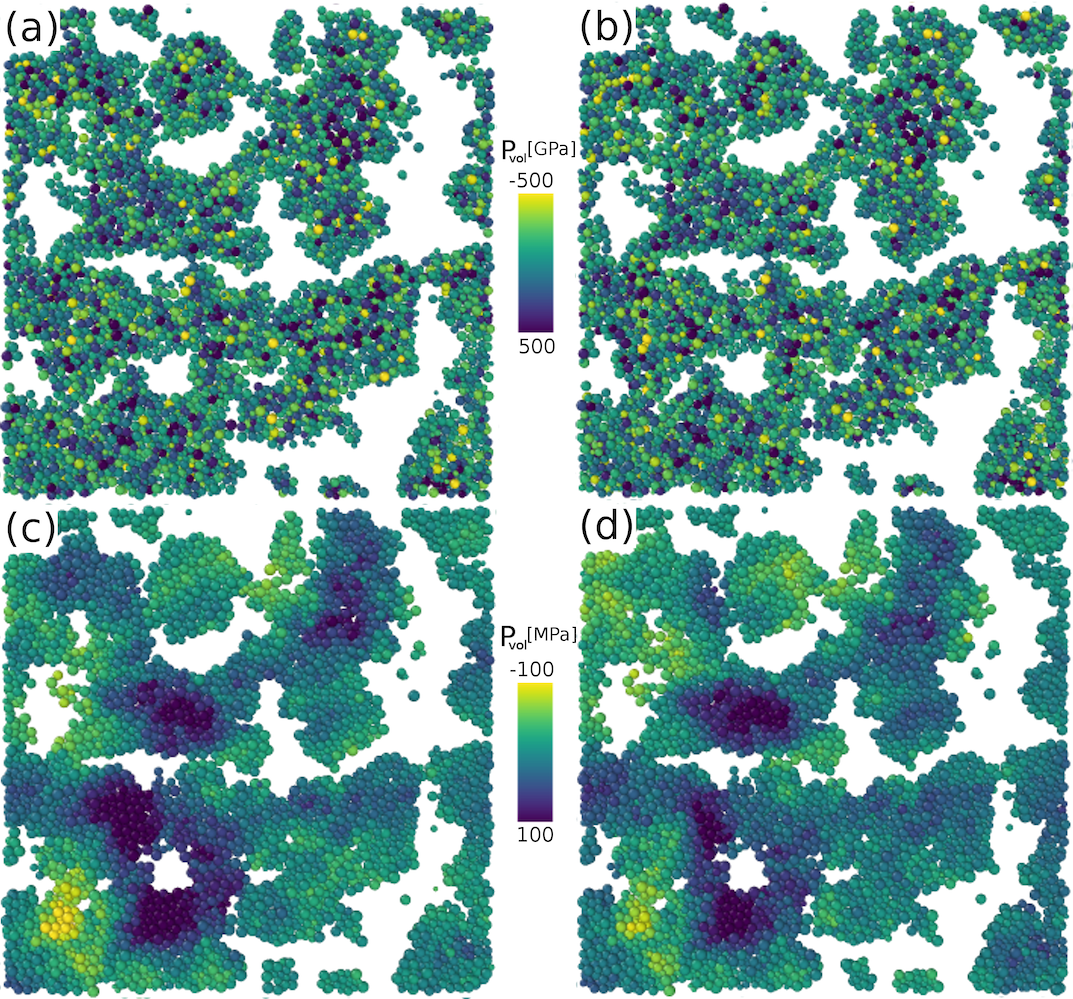}
\caption{
\label{fig:SI-homo-solid}
The solid stresses for (a) aged and (b) capillary aged samples. (c) (d) show the same solid stresses but homogenized at R=35~nm.
}
\end{figure}

\subsection{Stress homogenization and deformation analysis}
Under the assumption of mechanical equilibrium, the Love-Weber stress takes the same mathematical form as the Virial stress\citep{tsai1979virial}: 
\begin{equation}
\tensor{\bm\sigma}_{ij} = \frac{1}{V} \sum_{l\in \partial V} F^{cap,l}_{i}x^l_j
\label{eqn:love-weber}
\end{equation}
where one sums over particles indexed by $l$ on the boundary $\partial V$ of the homogenization volume element V, which is a sphere of radius $R$ centered on each particle. $x^l_j$ is the jth coordinate of particle $l$, $F^{cap, l}_i$ the ith component of capillary force on particle $l$
Fig.\ref{fig:SI-displacementmag} (a)(b) show the displacement fields at $RH=31\%$ and $RH=93\%$. Throughout this work all deformation analysis were done with a cut-off radius equal 2 times of average nano-particle radius. Homogeneous background is always subtracted when calculating the strain fields.
As $RH$ increases, the distribution of $D_{min}^2$ slightly narrows so that more particles experience very small non-affine deformations, as shown in Fig.\ref{fig:SI-displacementmag}(d). We show a cross section view of the non-affine deformations at RH=31\% and 93\% in Fig.\ref{fig:SI-Dmin2-crosssection}.

The homogenized capillary stress field is calculated according to Eqn.(1) in main text. A Love stress tensor is obtained for each particle for a given sphere radius. The distribution of liquid stress is shown in Fig.\ref{fig:SI-homo-pdf} (d), together with the solid stress PDFs (a-c). The solid stresses in aged and capillary aged samples are contrasted in Fig.\ref{fig:SI-homo-solid}(a)(b), together with their homogenized version in (c)(d).

\acknow{The authors would like to thank E. Masoero (U Newcastle), S. Yip and P. Cao (MIT)
for helpful discussions. This work has been supported by the Concrete Sustainability Hub at
MIT, A*MIDEX (the Aix-Marseille University  foundation) and CNRS. K.I. acknowledges the support of  the CEMCAP CNRS-Momentum program.}

\showacknow

\pnasbreak

\bibliography{cementdrying.bib}

\end{document}